\pdfoutput=1 
\documentclass{article}

\usepackage{epsf,hyperref}
\usepackage{amssymb,ComplexSystems} %comment_A
\usepackage{balance}
\usepackage{graphicx}
\usepackage{changepage}
\usepackage{caption}
\usepackage{subcaption}
\usepackage{multicol}
\usepackage{adjustbox}
\usepackage{comment}
\usepackage{calc}
\usepackage{times}
\usepackage{url}
\usepackage{hyperref}
\usepackage[htt]{hyphenat}

\begin{document}

\title{Causal Paths in Temporal Networks of Face-to-Face Human Interactions% 
% Use \\ to indicate line breaks in titles longer than about 
% 55 characters. 
%
}

\author{\authname{Agostino Funel}\\[2pt] 
% Use \\[2pt] to end the line and add space between author name and affiliation. 
\authadd{ENEA - Energy Technologies Department, ICT HPC Lab}\\
\authadd{Via E. Fermi, 1}\\
\authadd{80055 Portici (Naples), Italy}\\
\and
% For extra space, precede the second set of authors with \and.
%\authname{Second Author}\\
% Each author name goes on its own line.
%\authname{Third Author}\\[2pt]
%\authadd{Group, Laboratory}\\
%\authadd{Address}\\
%\authadd{City, State ZIP/Zone, Country}
% Do not use a ``.'' at the end of any line in the address. 
}

% The following specifies the running headings 
%
% Each running heading should be less than about 50 characters long. 
% If necessary, give a shortened version of the title. 
%
% Use initials for first and second names. If all author names do not fit, truncate the 
% list and end with ``et al.''.
%\markboth{Complex Systems} %commented_by_me
%{Sample Paper for Complex Systems} %commented_by_me

\maketitle
% End title section

\begin{abstract}
In a temporal network causal paths are characterized by the fact that links from a source to a target must respect the chronological order. In this article we study the causal paths structure in temporal networks of human face to face interactions in different social contexts. In a static network paths are transitive i.e. the existence of a link from $a$ to $b$ and from $b$ to $c$ implies the existence of a path from $a$ to $c$ via $b$. In a temporal network the chronological constraint introduces time correlations that affects transitivity. A probabilistic model based on higher order Markov chains shows that correlations that can invalidate transitivity are present only when the time gap between consecutive events is larger than the average value and are negligible below such a value. The comparison between the densities of the temporal and static accessibility matrices shows that the static representation can be used with good approximation. Moreover, we quantify the extent of the causally connected region of the networks over time.
%Data show that time correlations that can invalidate transitivity are important only for inter-events time windows larger than the average value meaning that only when the time gap beteen events outside these time intervals the static representation is inadequate. For each of the examined networks, we quantify the exention of causally connected regions over time, and give the distribution of shortest path durations.
\end{abstract}


% The text of the paper follows. All of the text should be in the same file. 
% Use separate files for large tabular material and graphics.

\section{Introduction}
\label{intro}
% \label is a hyperlink target for cross-referencing to this section using \ref{intro} (optional).
Many real systems composed of elements connected to each other can be described as networks. Notable examples are the Internet, transport networks, power grids. These technological and infrastructural networks can be considered static over time scales of many human daily activities because their structure does not change. On the contrary, human interactions can be thought of as events occurring over time. Product ratings, emails, social networks are examples of temporal networks~\cite{HOLME201297} because the topology changes over time. The set of interactions between any two nodes of the network can be described as a set of links to each of which are associated the nodes that interacted and the time  when the interaction occurred. Interactions involving many nodes can be treated considering all links which traversed them during the temporal interval in which the interactions occurred. The study of social dynamics is becoming increasingly important thanks to the availability of large datasets collected with technologies capable of monitoring human activities at different levels of space-temporal scales. Consider, for example, mobile  calls, instant messaging, digital traces~\cite{Monsivais2017, Leskovec2008, Badawy2018}. One might think that data analysis sheds light on human behavior~\cite{Tremblay2013,Lu2011}. Many aspects of human interactions can be analyzed considering the static time-aggregated network in which a link between any two nodes exists if they interacted at least once. Additional statistical information can be obtained by associating to each link a weight which is the number of times the link has been active during the observation period.  However, the (weighted) static representation may not be adequate to describe the underlying dynamics stemming from human interactions which can often be described as a diffusive process over the social network or by means of models where individuals are agents who change their state over time according to specific rules~\cite{ZHANG20161, Zhan2009, Opinion2017,Deffuant2000}. Moreover, in these cases events obey the principle of causality. When studying human interactions it is crucial to specify whether they are direct or mediated. With the advent of wearable devices it is now possible to study human interactions at the level of face to face proximity. In this context we can represent personal encounters as a collection of pathways which traverse all the people who interacted. In this article we study the causal paths structure in temporal networks of face to face human interactions in different social contexts. When causality is involved one must take into account the temporal order in which the links of a path occur. 
 The ordering of the links introduces correlations which may invalidate results based on a static time-aggregated network~\cite{PhysRevLett.110.198701,Scholtes2004,Rosvall2014}. For temporal networks there are fewer well established analysis techniques than for static representations. We investigate whether and to what extent the static representation is justified for the examined networks using two different probabilistic models based on higher order Markov chains and accessibility matrices. We also quantify the extent of the causally connected region of the networks during the period of observation.

The article is organized as follows: in Sec.~\ref{probmodels} we introduce the notation and the probabilistic models; we describe the datasets in Sec.~\ref{datasets} and report the results in Sec.~\ref{results}. We make the final conclusions in Sec.~\ref{conclusion}.

\section{Probabilistic models}
\label{probmodels}
In this section we introduce the notation and the probabilistic models used to analyze the datasets. A temporal network $G_{T} = (V, E_{T})$ can be defined as a set of nodes (or vertices) $v_i \in V$ and edges (or links) $E_{T} = V \times V \times [0, T]$ where $[0, T]$ is the temporal interval of observation. The interaction between two nodes $v_1$ and $v_2$  which starts at time  $t$ and has duration $\omega$ is an event that can be represented as a time-stamped link $(v_1,v_2,t, \omega)$. In the case of the examined datasets we can consider the time as a discrete variable. Events whose duration is less than the temporal resolution $\epsilon$ of the data acquisition system can not be observed. In all other cases $\omega$ can be divided into $n$ elementary time steps $\omega = n  \sigma$ ($\sigma \geq \epsilon$) and the interaction $(v_1,v_2,t, \omega)$ can be replaced by the sequence of "instantaneous" events $(v_1,v_2,t_i)$ where $t_i = t + i \sigma \leq \omega$ and $i = 0,...,n$. A causal path between a source $v_0$ and a target $v_l$ is any sequence $(v_0, v_1, t_1),(v_1, v_2, t_2),...,(v_{l-2},v_{l-1},t_{l-1}),(v_{l-1},v_l,t_l)$ where $t_1 < t_2 <...< t_{l-1} < t_l$. The length $l$ of the path is the number of links which traverse its nodes. A single node is a path of length zero. It is sometimes useful to impose the constraint $0 < t_{i+1} - t_i \leq \delta$ between consecutive events to select only causal paths that contribute to dynamical processes whose characteristic time scale is $\delta$. For example, we may be interested in studying the propagation of information during conversations of average duration $\delta$ by analyzing data collected during a period $T \gg \delta$.

\subsection{Higher order Markov chains}
\label{markov}
We now briefly present the higher order Markov chains probabilistic model used to analyze the datasets. The model has been proposed in~\cite{Scholtes2017} and the reader is encouraged to read the article for more details. 
In the static time-aggregated network representation the existence of the links $(v_{i-1},v_i)$ and $(v_i,v_{i+1})$ implies the existence of the path $(v_{i-1},v_i,v_{i+1})$ then we would be led to think that $v_{i-1}$ can influence $v_{i+1}$ via $v_i$. However, this transitivity may be invalidated in the temporal network by time correlations because it exists only if $(v_{i-1},v_i)$ occurred before $(v_i,v_{i+1})$. For any given $\delta$ let $S = \{p_{(1)},...,p_{(N)}\}$ be the collection of all causal paths which satisfy the condition $0 < t_{i+1} - t_i \leq \delta$. A causal path $p_l$ of length $l$ can be considered a sequence of transitions over random variables $(v_0 \rightarrow v_1 \rightarrow ... \rightarrow v_l)$. In this model one define a discrete time Markov chain of order $k$ and assume that for a vertex $v_i$ of the path the probability to reach it depends only on the $k$ previously traversed vertices $P(v_i|v_0 \rightarrow ... \rightarrow v_{i-1}) = P(v_i|v_{i-k} \rightarrow ... \rightarrow v_{i-1})$. For a given order $k$ a maximum likelihood estimation is performed to find the transition probabilities $P^{(k)}:= P(v_i|v_{i-k} \rightarrow ... \rightarrow v_{i-1})$. It is found that the probabilities that maximize the likelihood function can be calculated from the relative frequencies of sub-paths of length $k$ in $S$. The order $k = 0$ only generates vertices of the network. For $k=1$ only sub-paths of length $1$ are considered and the model reproduces the topology of the static weighted network where the weight of each edge is its frequency. For $k > 1$ the model captures correlations of longer paths which can not be evaluated on the basis of the topology of the underlying static network. To a $k$-order model is associated a $k$-order aggregated network $G^{(k)}$~\cite{Scholtes2015}. Each node of $G^{(k)}$ is a $k$-tuple $v^{(k)}=(v_1,...,v_k)$ and a link $(v^{(k)},w^{(k)})$ exists between two nodes if $v_{i+1} = w_i$ for $i = 1,...,k-1$. A $k$-order Markov chain is treated as a first order Markov chain over nodes of $G^{(k)}$. To find which is the higher order network abstraction that best models the observed paths a procedure which combines multiple layers of Markov chains is used. For a given path $p_l$ the transition probability of the multi layer model of order $K$ is $\mathcal{P}^{(K)}(v_0 \rightarrow ... \rightarrow v_l) = \prod_{k=1}^{K-1}P^{(k)}(v_k|v_0 \rightarrow ... \rightarrow v_{k-1}) \prod_{i=K}^{l} P^{(K)}(v_i|v_{i-K} \rightarrow ... \rightarrow v_{i-1})$. The likelihood function of the multi layer order $K$ model is $\mathcal{L}_K = \prod_{j=1}^{N} \mathcal{P}^{(K)}(p_{(j)})$ and the optimal order $K_{opt}$ is found by progressively evaluating the likelihood ratio $\mathcal{L}_K/\mathcal{L}_{K+1}$ of models having consecutive orders. If $K_{opt} = 1$ correlations preserve the transitivity while it is broken if $K_{opt} > 1$. Thus in the first case the static time-aggregated network representation is justified to analyze data. We note that the set of observed causal paths $S$ depends on the inter-events time gap $\delta$, therefore $K_{opt}$ also depends on it.

\subsection{Accessibility matrices}
\label{timeunfoldaccmtx}
This model has been proposed in~\cite{Lentz2013} and the reader is advised to read the article for the details. In a static network the accessibility matrix provides a macroscopic view of the connections. Its elements are $1$ if the nodes corresponding to the indices are connected by a path of any length, and zero otherwise. In the static case the adjacency matrix $\mathbf{A}$ of the network is $\mathbf{A}_{ij} = 1$ if nodes $(i,j)$ are connected by an edge, $\mathbf{A}_{ij} = 0$ otherwise. The elements of the matrix $\mathbf{A}^{n}$ are the number of paths of length $n$ connecting the nodes. We define the binary operator $\mathcal{U}$ which acts on matrices in the following way: $\mathcal{U}(\mathbf{M}_{ij})= 1$ if $\mathbf{M}_{ij} \neq 0$ and $\mathcal{U}(\mathbf{M}_{ij})= 0$ if $\mathbf{M}_{ij} = 0$. The accessibility matrix is $\mathbf{P}_N =  \mathcal{U}(\sum_{i=1}^{N}\mathbf{A}^i)$ where $N$ is the number of nodes.  Since the maximum distance is the diameter $D$ of the network, the process of summing up saturates and one has $\mathbf{P}_N = \mathbf{P}_D$.  The density of a $N \times N$ matrix $\mathbf{M}$ is $\rho(\mathbf{M}) = \mbox{nnz}(\mathbf{M})/N^2$, where $\mbox{nnz}(\mathbf{M})$ is the number of nonzero elements. The probability to find a shortest path of length $l \leq n$ between two randomly chosen nodes is $P(l \leq n) = \rho(\mathbf{P}_n)$. For a disconnected network $\rho(\mathbf{P}_D) < 1$ because the number of nonzero entries of the accessibility matrices of the disjoint components is less than $N^2$. A temporal network can be represented by a sequence of chronologically ordered adjacency matrices $\mathcal{A} =\{ \mathbf{A}_{t_1},\mathbf{A}_{t_2}, ...,\mathbf{A}_{t_T} \}$. Causality is guaranteed by the condition $t_1 < t_2 < ... < t_T = T$. The accessibility matrix of the temporal network is $\mathbf{\mathcal{P}}_T =  \mathcal{U}(\prod_{i=1}^{T}(\mathbf{1} + \mathbf{A}_{t_i}))$. The density of the accessibility matrix $\rho(\mathbf{\mathcal{P}}_n)$ gives the fraction of the network causally connected for $t \leq t_n$. The model provides a criterion to quantify the goodness of the static time-aggregated approximation for a temporal network by evaluating the causal fidelity index $0 \leq \gamma = \rho(\mathbf{\mathcal{P}}_T) / \rho(\mathbf{P}_N) \leq 1$. Low values of $\gamma$ indicate bad approximation meaning that the majority of paths in the static representation do not correspond to causal paths in the temporal network. Instead, high values of $\gamma$ indicate good approximation.

\section{Datasets}
\label{datasets}
We use publicly available datasets provided by the SocioPatterns~\cite{sociopatterns} collaboration. The data acquisition system is based on a sensing platform which uses wearable badges equipped with Radio Frequency Identification (RFID) devices. Contacts data is gathered at the level of face to face proximity ($\sim$1.5 meter) with a temporal resolution of 20 seconds~\cite{rfid}. We consider different social contexts. \textbf{High School}: contacts between students of five classes in a high school in Marseilles (France) collected during 7 days in November 2012~\cite{highschool}; \textbf{Hospital}: contacts between patients and healthcare personnel in a hospital ward in Lyon (France) collected during five days in December 2010~\cite{hospital}; \textbf{HT09}: encounters between people who attended the ACM Hypertext Conference held in Turin (Italy) in 2009~\cite{ht09inf}; \textbf{Infectious}: encounters between people during the Infectious exhibition event held at the Science Gallery in Dublin (Ireland) from April 17th to July 17th in 2009~\cite{ht09inf}; \textbf{SFHH}: interactions between participants to a scientific conference in Nice (France), June 4-5, 2009~\cite{sfhhwp2};  \textbf{Workplace}: contacts between individuals in an office building in France in 2015~\cite{sfhhwp2}.

Since all measurements of our interest are invariant under time translation, to better compare the temporal networks for all of them we make sure that events start at $t = 0$. Table~\ref{tab_datasets} shows the size and temporal information of the datasets.

%\begin{comment}
\begin{table*}[h!]
\begin{center}
\begin{adjustbox}{max width=\textwidth}
\begin{tabular}{|c|c|c|c|c|c|}
\hline
Network &       Nodes   &       Time-stamped links      &       Observation duration (sec)      &       $\delta_{avg}$ (sec)    &       $\delta_{max}$ (sec)    \\
\hline
\hline
High School     &       180     & 45047 &       729500  &       64.72   &       220280   \\
\hline
Hospital        &       75      & 32424 &       347500  &       36.76   &       26980   \\
\hline
HT09            &       113     & 20818 &       212340  &       40.48   &       28900   \\
\hline
Infectious      &       10972   & 415912 &      6946340 &       90.28   &       152980  \\
\hline
SFHH            &       403     & 70261 &       114300  &       32.58   &       38320   \\
\hline
Workplace       &       217     & 78249 &       993540  &       53.74   &       218600  \\
\hline
\end{tabular}
\end{adjustbox}
\end{center}
\caption{Human face to face interaction temporal networks analyzed in this work. The minimum inter-events time $\delta_{min}$ is equal to 20 sec which is the time resolution of the data acquisition system. The average (maximum) time gap between consecutive events is $\delta_{avg}$ ($\delta_{max}$).}
\label{tab_datasets}
\end{table*}
%\end{adjustwidth}
%\end{comment}

The software libraries we use in this work are: \texttt{tacoma}~\footnote{https://github.com/benmaier/tacoma} to study the edge activity; \texttt{pathpy}~\footnote{https://github.com/IngoScholtes/pathpy} to study the causal paths statistics and perform the  multi layer Markov chains analysis; and the \texttt{TemporalNetworkAccessibility}~\footnote{https://github.com/hartmutlentz/TemporalNetworkAccessibility} classes to compute the accessibility matrices.

\section{Results}
\label{results}

\begin{figure}[htbp!]
  \begin{center}
    \includegraphics[width=\textwidth]{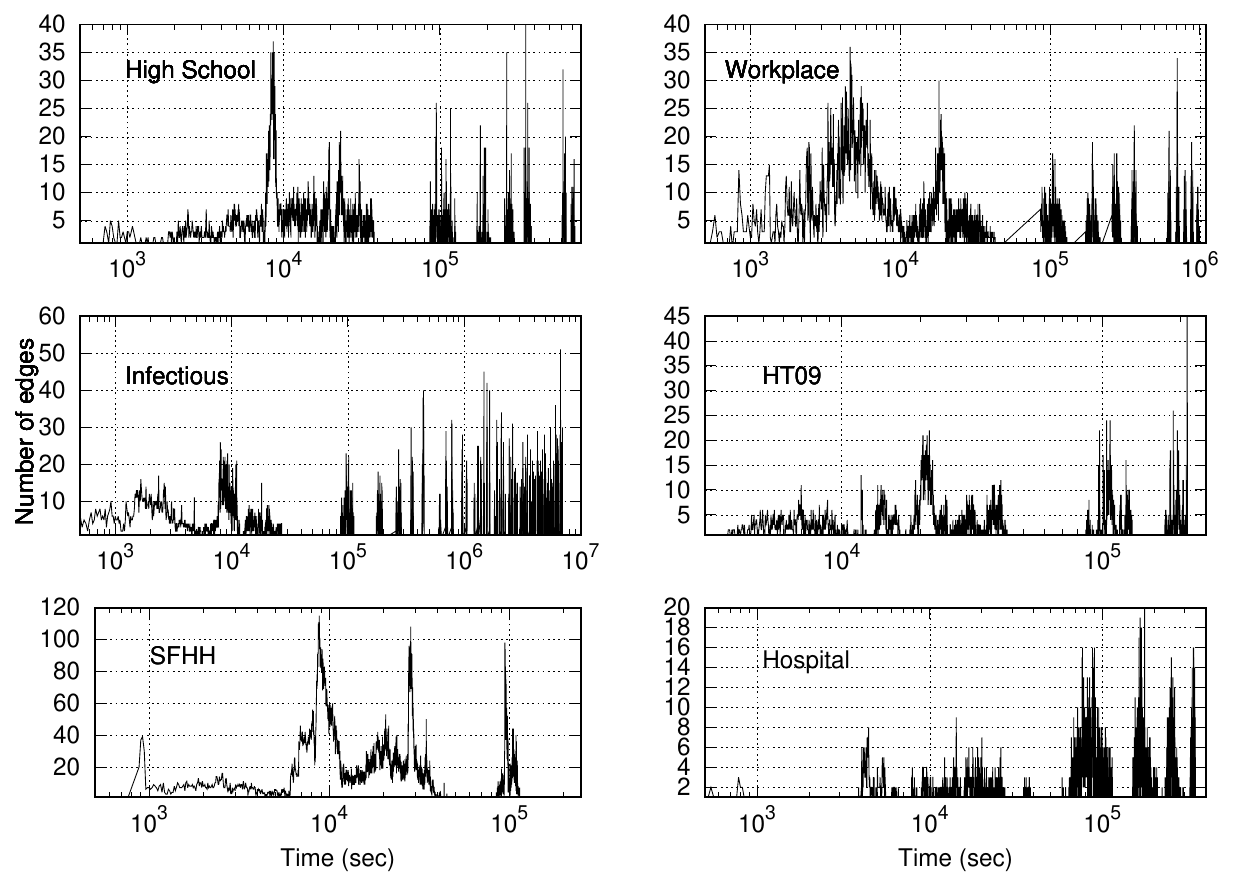}
\caption{Number of active edges during the observation period. }
\label{fig_edgecount}
  \end{center}
\end{figure}

\begin{figure}[htbp!]
  \begin{center}
    \includegraphics[width=\textwidth]{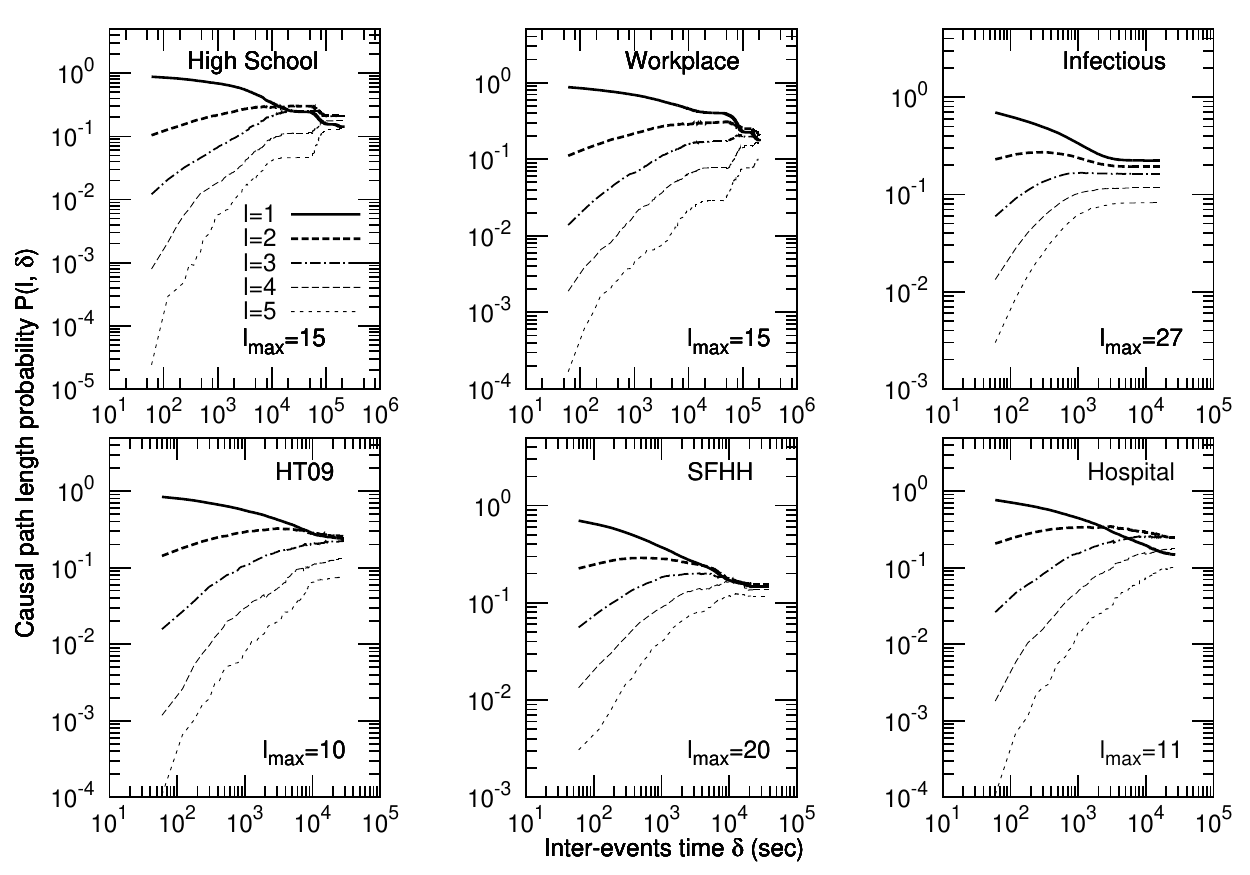}
\caption{Probability distributions $P(l, \delta)$ of causal paths of length $l$ as a function of the inter-events time $\delta$. The maximum observed path length is $l_{max}$.}
\label{fig_plpd}
  \end{center}
\end{figure}

\begin{figure}[htbp!]
  \begin{center}
    \includegraphics[width=\textwidth]{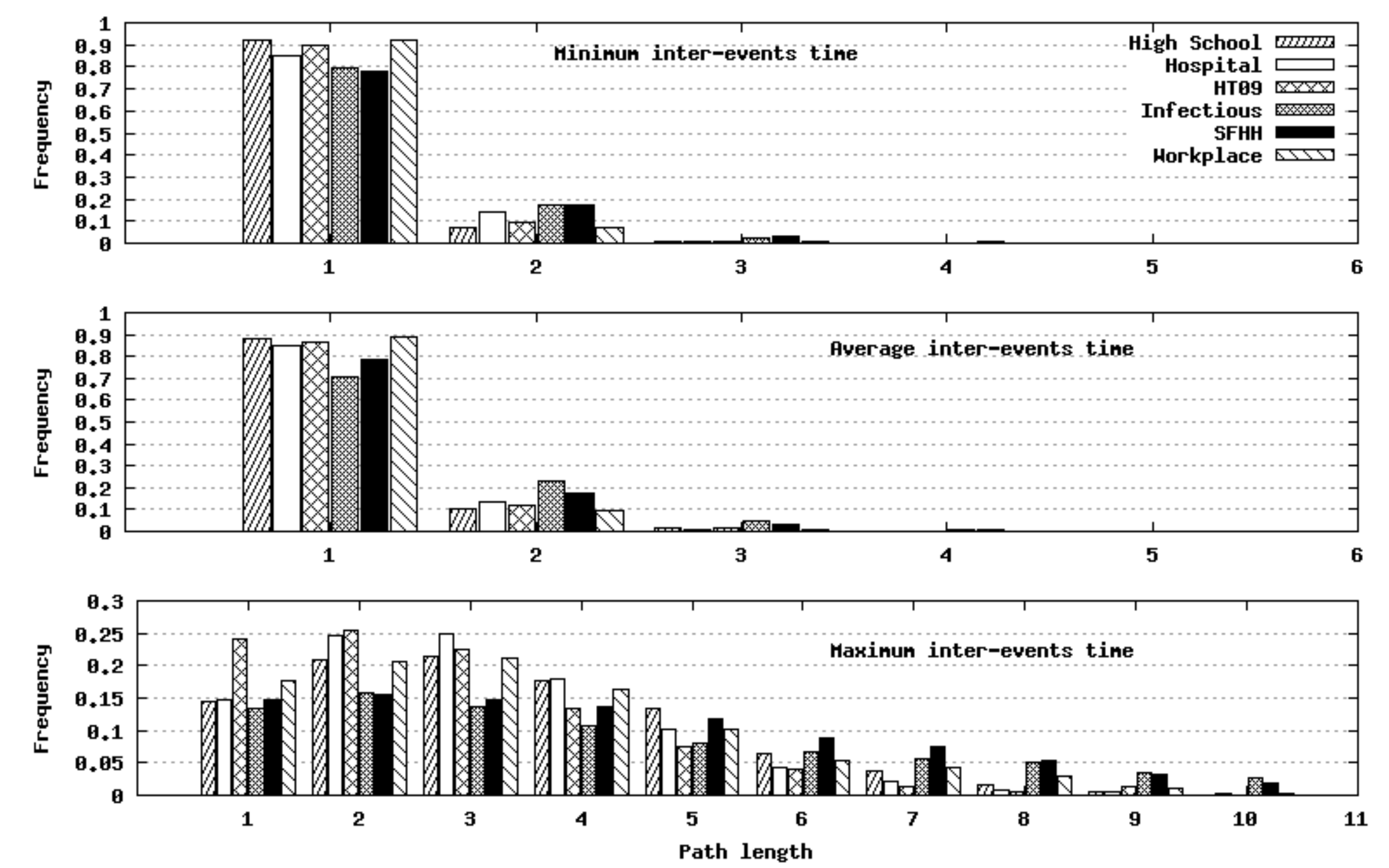}
\caption{Probability distributions of causal paths as a function of the length for the inter-events times $\delta = \delta_{min}$ (top), $\delta = \delta_{avg}$ (middle) and $\delta = \delta_{max}$ (bottom). }
\label{fig_hist_plpd}
  \end{center}
\end{figure}

\begin{figure}[htbp!]
  \begin{center}
    \includegraphics[width=\textwidth]{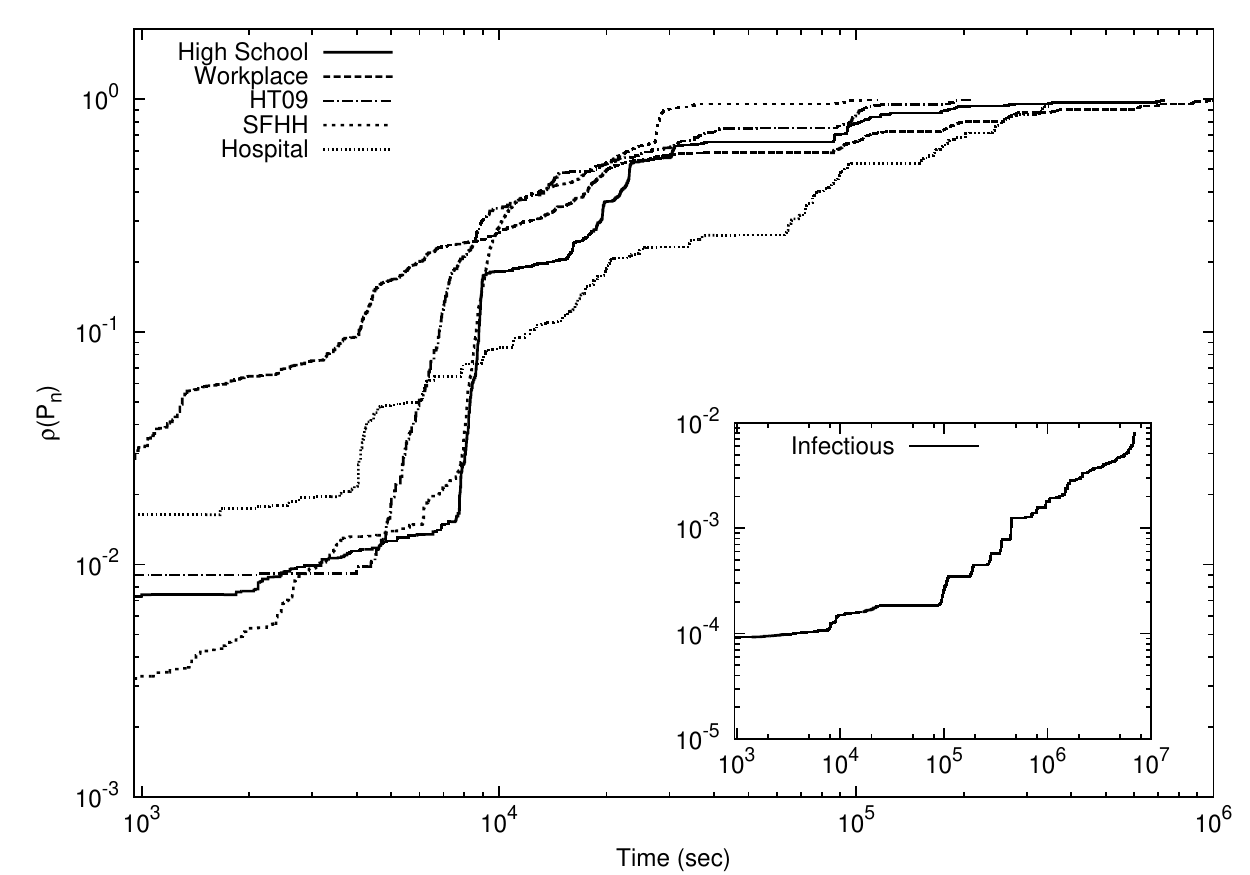}
\caption{Densities $\rho(\mathcal{P}_n)$ of the accessibility matrices of the temporal networks as a function of the time.}
\label{fig_path_density}
  \end{center}
\end{figure}

The overall edge activity is plotted in Fig.~\ref{fig_edgecount} for all networks. Inactivity periods during which data was not collected are clearly shown. The total number of causal paths depends on the inter-events time window $\delta$. Looking at Tab.~\ref{tab_datasets} we observe that for all networks the minimum and average time gap between consecutive events $\delta_{min}$ and $\delta_{avg}$ is of the order of tens of seconds while the maximum $\delta_{max}$ is two or three order of magnitude larger. To sample the interval $[\delta_{min}, \delta_{max}]$ we divide it in sub-intervals of equal width. We choose the sampling width $\mu = 1$ minute and collect the path statistics at $\delta_i = \delta_{min} + i \mu \leq \delta_{max} \quad i = 0,1,...,m$. Fig.~\ref{fig_plpd} shows the probability distributions $P(l, \delta)$ of causal paths of length $l$ as a function of $\delta$. It is interesting to note that the distributions $P(l, \delta)$ have the same trend for all social contexts. The probability $P(1,\delta)$ decreases with $\delta$ while $P(l > 1,\delta)$ increases in almost the whole range $[\delta_{min}, \delta_{max}]$. In the figure are plotted only the probability distributions for paths of length $l \leq 5$ for better readability. The distributions $P(l > 5, \delta)$ all have the same shape of $P(l > 1, \delta)$ and $P(l + 1, \delta) < P(l, \delta)$. In the interval $[\delta_{min}, \delta_{avg}]$ the percentage of paths of length $l = 1$ varies in the range $\sim$[70--90]\%, that of paths of length $l = 2$ in the range $\sim$[7--23]\%, and paths of length $l \geq 3$ are rare. In the limit $\delta \rightarrow \delta_{max}$ the contribution of paths of length $ 3 \leq l \leq 10$ to the statistics is significantly higher. However, even in this limit paths of length $l > 10$ are rare. Fig.~\ref{fig_hist_plpd} shows for $\delta = \delta_{min}, \delta_{avg}$ and $\delta_{max}$ the probabilities of causal paths of length $l \leq 10$. Time correlations in causal paths come into play only when $l \geq 2$ and their importance is greater the greater is the length. Therefore, we expect that transitivity is not lost around $\delta_{avg}$ but only for $\delta > \delta_{avg}$. This is confirmed by the higher order multi layer Markov chains model. It is difficult to predict the exact value of $\delta$ for which a first order network abstraction is inadequate. However, we argue that since for the majority of the links of a causal path the time gap between consecutive events is around $\delta_{avg}$, the first order approximation could be acceptable and the static time-aggregated network model could be used with a good degree of confidence. We thus compute the causal fidelity index $\gamma$ to quantify the goodness of the static representation. The results of both models are reported in Tab.~\ref{tab_koptgamma}. A part from the "Infectious" network for which $\gamma = 0.57$, we observe a value of $\gamma \approx 1$ in all other cases. This indicate that regardless of the social context the static time-aggregate network representation is a good approximation of the causal paths structure and may be trusted to describe dynamical processes whose time scale does not differ to much from the average inter-events time gap. The density of the accessibility matrix $\rho(\mathcal{P}_n)$ is plotted in Fig.~\ref{fig_path_density} for all networks. It gives information on the extent of the causally connected region of the network over time.

\begin{table*}[htbp!]
\begin{center}
\begin{adjustbox}{max width=\textwidth}
\begin{tabular}{|c|ccc|c|}
\hline
Network &       \multicolumn{3}{ |c| }{$K_{opt}$}  & $\gamma$        \\
        &$\delta_{min}$   &       $\delta_{avg}$      &    $\delta_{max}$    &      \\
\hline
\hline
High School     &       1     & 1 &       2  &       0.97      \\
\hline
Hospital        &       1     & 1 &       2  &       0.93      \\
\hline
HT09            &       1     & 1 &       2  &       0.99      \\
\hline
SFHH            &       1     & 1 &       2  &       0.99      \\
\hline
Workplace       &       1     & 1 &       2  &       0.99     \\
\hline
Infectious      &       1     & 1 &       2 &        0.57     \\
\hline
\end{tabular}
\end{adjustbox}
\end{center}
\caption{The table shows, for all examined temporal networks, the optimal maximum order $K_{opt}$ provided by the multi layer Markov chains model evaluated for minimum, average and maximum inter-events times, and the causal fidelity index $\gamma$.  }
\label{tab_koptgamma}
\end{table*}
%\end{adjustwidth}

\begin{table*}[htbp!]
\begin{center}
\begin{adjustbox}{max width=\textwidth}
\begin{tabular}{|c|c|c|c|c|}
\hline
        &       \multicolumn{4}{ |c| }{\small Time (\%$T$) taken to reach}\\
	&	\multicolumn{4}{ |c| }{\small the following percentages}  \\
Network        &    \multicolumn{4}{ |c| }{\small of the network causally connected}  \\
        &              25\%  &        50\%    &      75\%   &    90\%    \\
\hline
\hline
High School     &       2.4     & 3.2 &      13.0   & 24.8    \\
\hline
Hospital        &       10.6     & 26.7 &       70.0    & 95.3  \\
\hline
HT09            &       4.0     & 8.9 &   20.0        & 48.9 \\
\hline
SFHH            &       8.4     & 15.8 &       24.2   & 25.6    \\
\hline
Workplace       &       0.88     & 2.0 &       18.0   & 37.3    \\
\hline
Infectious	& \multicolumn{4}{ |c| }{\small $\sim$1\% causally connected at the}  \\ 
		& \multicolumn{4}{ |c| }{\small end of the observation period $T$}  \\
\hline
\end{tabular}
\end{adjustbox}
\end{center}
%%%%%%%%%%%%%%%%%%%%%%%%
\caption{The table shows the percentage of the observation period $T$ taken to have one quarter, half, three-quarters and 90\%  of the network causally connected. The "Infectious" network is only $\sim$1\% causally connected at the end of $T$.}
\label{tab_tccr}
\end{table*}
%\end{adjustwidth}

Tab.~\ref{tab_tccr} shows the extent of the causally connected region of the networks during the observation period $T$. The "Infectious" network is $\sim$1\% causally connected at the end of the observation period. Its static time-aggregated network is not connected. This means that there are components which never were connected during the observation period. These components contain isolated individuals or disjoint groups of individuals. Note that the analyzed data of the "Infectious" network were collected from April 17th to July 17th 2009 therefore it is possible that people who participated to the exhibition event on different days never met.
We observe that the "High School" and "SFHH" networks are 90\% causally connected after about one quarter of the observation period, meaning a high level of mobility and interaction between individuals. We may think that students of a class had frequent relationships not only between them but also with students of other classes, for example during recess time after courses, and people who attended the scientific conference frequently met to exchange opinions. The "Hospital" network becomes 90\%  causally connected only at the end of the observation period and the extent of the network region connected by causal paths increases quite slowly. The "HT09" network becomes 90\% causally connected after $\sim$50\% of the observation period which is a sign that the majority of people had a face to face contact already during the first half of the conference duration time. The "Workplace" network is 25\% causally connected just after $\sim$1\% of the observation period and becomes 90\% causally connected after $\sim$37\%. This is an indication that the employees who worked in the office building had frequent relationships probably because their activities required strong team collaboration.

\section{Conclusion}
\label{conclusion}
We studied the causal paths structure in temporal networks of human face to face proximity interactions in different social contexts. The importance of causal paths lies in the fact that they are the underlying background on top of which diffusive processes evolve. Temporal networks are often analyzed by applying well established algorithms and techniques to the static time-aggregated representation. Although this approach provides information on global properties, it implicitly assumes link transitivity and does not capture time correlations. Therefore, it  may lead to misleading results in the analysis of causal paths. To quantify time correlations and the goodness of the static approximation we applied two independent models to the datasets, one based on multi layer higher order Markov chains, and the other on the unfolding of the accessibility matrices. We found that for all examined networks the probability distributions of causal paths as a function of the waiting time $\delta$ between consecutive events have the same shape.  The number of causal paths of length one decreases with $\delta$ and that of longer paths increases. The number of causal paths of length one considerably exceeds that of longer paths for $\delta$ less than or equal to the average value $\delta_{avg}$. Time correlations are present for paths of length $l \geq 2$. Their number increases with $\delta$ but exceeds that of path of length one only for $\delta \gg \delta_{avg}$. For all examined networks the  multi layer higher order Markov chains model shows that transitivity is not lost when  $\delta = \delta_{avg}$ indicating that for temporal scales of this order the static network representation may be a good approximation. We evaluated the ratio of the densities of the  temporal and static accessibility  matrices which provides an index $ 0 \leq \gamma \leq 1$. High values of $\gamma$ indicate that the static representation is a good approximation and low values bad approximation. Except for one of the examined network (exhibition event lasting about three months) for which $\gamma \sim 0.6$, for all other networks we found that $\gamma \sim 1$.  We also quantified the extent of the causally connected region of the networks during the observation period obtaining an overall view of the people mobility in the different social contexts.    

\section{Acknowledgments}
The computing resources and the related technical support used for this work have been provided by CRESCO/ENEAGR\-ID High Performance Computing infrastructure and its staff~\cite{eneagrid}. CRESCO/ENEAGRID High Performance Computing infrastructure is funded by ENEA, the Italian National Agency for New Technologies, Energy and Sustainable Economic Development and by Italian and European research programmes, see https://www.eneagrid.enea.it for information.

\bibliographystyle{unsrt}
\bibliography{human_contacts_pathaways}

\begin{thebibliography}{10}

\bibitem{HOLME201297}
Petter Holme and Jari Saram{\"a}ki.
\newblock Temporal networks.
\newblock {\em Physics Reports}, 519(3):97 -- 125, 2012.
\newblock Temporal Networks.

\bibitem{Monsivais2017}
Daniel Monsivais, Asim Ghosh, Kunal Bhattacharya, Robin I.~M. Dunbar, and Kimmo
  Kaski.
\newblock {Tracking urban human activity from mobile phone calling patterns}.
\newblock {\em PLOS Computational Biology}, 13(11):1--16, 11 2017.

\bibitem{Leskovec2008}
Jure Leskovec and Eric Horvitz.
\newblock {Planetary-Scale Views on a Large Instant-Messaging Network}.
\newblock In {\em Proceedings of the 17th International Conference on World
  Wide Web}, WWW ’08, page 915–924, New York, NY, USA, 2008. Association
  for Computing Machinery.

\bibitem{Badawy2018}
Adam Badawy, Emilio Ferrara, and Kristina Lerman.
\newblock {Analyzing the Digital Traces of Political Manipulation: The 2016
  Russian Interference Twitter Campaign}.
\newblock {\em CoRR}, abs/1802.04291, 2018.

\bibitem{Tremblay2013}
Nicolas Tremblay, Alain Barrat, Cary Forest, Mark Nornberg, Jean-Fran\c{c}ois
  Pinton, and Pierre Borgnat.
\newblock {Bootstrapping under constraint for the assessment of group behavior
  in human contact networks}.
\newblock {\em Phys. Rev. E}, 88:052812, Nov 2013.

\bibitem{Lu2011}
X.~Lu and C.~Brelsford.
\newblock {Network Structure and Community Evolution on Twitter: Human Behavior
  Change in Response to the 2011 Japanese Earthquake and Tsunami.}
\newblock {\em Sci. Rep.}, 4, 2015.

\bibitem{ZHANG20161}
Zi-Ke Zhang, Chuang Liu, Xiu-Xiu Zhan, Xin Lu, Chu-Xu Zhang, and Yi-Cheng
  Zhang.
\newblock Dynamics of information diffusion and its applications on complex
  networks.
\newblock {\em Physics Reports}, 651:1 -- 34, 2016.

\bibitem{Zhan2009}
X.~Zhan, A.~Hanjalic, and H.~Wang.
\newblock {Information diffusion backbones in temporal networks.}
\newblock {\em Sci. Rep.}, 9, 2019.

\bibitem{Opinion2017}
Alina Sirbu, Vittorio Loreto, Vito D.~P. Servedio, and Francesca Tria.
\newblock {\em {Opinion dynamics: models, extensions and external effects}},
  page 363{\textendash}401.
\newblock Springer, 2017.

\bibitem{Deffuant2000}
Guillaume Deffuant, David Neau, Frédéric Amblard, and Gérard Weisbuch.
\newblock {Mixing Beliefs Among Interacting Agents}.
\newblock {\em Advances in Complex Systems}, 3:87--98, 01 2000.

\bibitem{PhysRevLett.110.198701}
Ren\'e Pfitzner, Ingo Scholtes, Antonios Garas, Claudio~J. Tessone, and Frank
  Schweitzer.
\newblock {Betweenness Preference: Quantifying Correlations in the Topological
  Dynamics of Temporal Networks}.
\newblock {\em Phys. Rev. Lett.}, 110:198701, May 2013.

\bibitem{Scholtes2004}
Ingo Scholtes, Nicolas Wider, René Pfitzner, Antonios Garas, Claudio Tessone,
  and Frank Schweitzer.
\newblock {Causality-Driven Slow-Down and Speed-Up of Diffusion in
  Non-Markovian Temporal Networks}.
\newblock {\em Nature communications}, 5:5024, 09 2014.

\bibitem{Rosvall2014}
Martin Rosvall, Alcides Viamontes~Esquivel, Andrea Lancichinetti, Jevin West,
  and Renaud Lambiotte.
\newblock {Memory in network flows and its effects on spreading dynamics and
  community detection}.
\newblock {\em Nature communications}, 5:4630, 08 2014.

\bibitem{Scholtes2017}
Ingo Scholtes.
\newblock {When is a Network a Network? Multi-Order Graphical Model Selection
  in Pathways and Temporal Networks}.
\newblock In {\em Proceedings of the 23rd ACM SIGKDD International Conference
  on Knowledge Discovery and Data Mining}, KDD ’17, page 1037–1046, New
  York, NY, USA, 2017. Association for Computing Machinery.

\bibitem{Scholtes2015}
Ingo Scholtes, Nicolas Wider, and Antonios Garas.
\newblock {Higher-Order Aggregate Networks in the Analysis of Temporal
  Networks: Path structures and centralities}.
\newblock {\em The European Physical Journal B}, 89, 08 2015.

\bibitem{Lentz2013}
Hartmut H.~K. Lentz, Thomas Selhorst, and Igor~M. Sokolov.
\newblock {Unfolding Accessibility Provides a Macroscopic Approach to Temporal
  Networks}.
\newblock {\em Phys. Rev. Lett.}, 110:118701, Mar 2013.

\bibitem{sociopatterns}
SocioPatterns collaboration.
\newblock \url{http://www.sociopatterns.org/}.

\bibitem{rfid}
Ciro Cattuto, Wouter {Van den Broeck}, Alain Barrat, Vittoria Colizza,
  {Jean-François} Pinton, and Alessandro Vespignani.
\newblock {Dynamics of Person-to-Person Interactions from Distributed {RFID}
  Sensor Networks}.
\newblock {\em PLOS ONE}, 5(7):e11596, 07 2010.

\bibitem{highschool}
Julie Fournet and Alain Barrat.
\newblock {Contact Patterns among High School Students}.
\newblock {\em PLoS ONE}, 9(9):e107878, 09 2014.

\bibitem{hospital}
Philippe Vanhems, Alain Barrat, Ciro Cattuto, Jean-Fran\c{o}is Pinton, Nagham
  Khanafer, Corinne R\'{e}gis, Byeul-a Kim, Brigitte Comte, and Nicolas Voirin.
\newblock {Estimating Potential Infection Transmission Routes in Hospital Wards
  Using Wearable Proximity Sensors}.
\newblock {\em PLoS ONE}, 8(9):e73970, 09 2013.

\bibitem{ht09inf}
Lorenzo Isella, Juliette Stehlé, Alain Barrat, Ciro Cattuto,
  {Jean-Fran\c{c}ois} Pinton, and Wouter {Van den Broeck}.
\newblock {What's in a Crowd? Analysis of Face-to-Face Behavioral Networks}.
\newblock {\em Journal of Theoretical Biology}, 271(1):166--180, 2011.

\bibitem{sfhhwp2}
Mathieu G{\'e}nois and Alain Barrat.
\newblock Can co-location be used as a proxy for face-to-face contacts?
\newblock {\em EPJ Data Science}, 7(1):11, May 2018.

\bibitem{eneagrid}
F.~Iannone et~al.
\newblock {CRESCO} {ENEA} {HPC} clusters: a working example of a multifabric
  {GPFS} {S}pectrum {S}cale layout.
\newblock {\em PROC. HPCS}, 2019.

\end{thebibliography}

\end{document}